\documentclass[prl, twocolumn,showpacs]{revtex4} 
\usepackage{amsmath}
\usepackage{amsfonts}
\usepackage{amssymb}
\usepackage{graphics}
\usepackage{graphicx}

\def\hom{Hong-Ou-Mandel }
\def\sbra#1{\langle #1\mid} 
\def\sket#1{|#1\rangle}
\def\d#1{#1^\dagger}

\begin{document}
\title{Quantum interference of ultrastable twin optical beams}
\author{Sheng Feng}
\author{Olivier Pfister}
\email[Corresponding author: ]{opfister@virginia.edu}
\affiliation{Department of Physics, University of Virginia, 382 McCormick Road, Charlottesville, VA 22904-4714, USA}
\begin{abstract}
We report the first measurement of the quantum phase-difference noise of an ultrastable nondegenerate optical parametric oscillator that emits twin beams classically phase-locked at exact frequency degeneracy. The measurement illustrates the property of a lossless balanced beam-splitter to convert number-difference squeezing into phase-difference squeezing and, thus, provides indirect evidence for Heisenberg-limited interferometry using twin beams. This experiment is a generalization of the Hong-Ou-Mandel interference effect for continuous variables and constitutes a milestone towards continuous-variable entanglement of bright, ultrastable nondegenerate beams.  
\end{abstract}
\maketitle 

Nonclassical interference of highly excited boson modes is of fundamental interest for ultra-precise physical measurements, such as Heisenberg-limited interferometry  \cite{hli} and spectroscopy \cite{meyer}, which find applications in ultimate-sensitivity measurements such as gravitational-wave detection \cite{gwd} and modern atomic clocks \cite{clairon}. Moreover, the quantum noise reduction (squeezing) at the heart of Heisenberg-limited measurements is connected to continuous-variable entanglement \cite{reid}, of interest for quantum information and communication \cite{furusawa}. Preliminary Heisenberg-limited interferometers with $N=2$ bosons have been realized with twin photons \cite{kuzmich} and trapped ions \cite{meyer}. Ultrasensitive interferometry, however, requires $N\gg 1$ and continuous-variable quantum optics and squeezed states are the tools of choice here. The use of Bose-Einstein condensates has been proposed \cite{bouyer} and progress has been made in this direction \cite{kas}. Recently, Silberhorn {\em {\em et al.}}\ made a beautiful demonstration of continuous-variable entanglement of picosecond-pulsed optical beams, by simultaneous squeezing of the number sum and of the phase difference \cite{silb}. For high-precision measurements, however, stable CW beams are preferable. One interesting system for this purpose is the ultrastable nondegenerate optical parametric oscillator (OPO), which emits intense twin beams. In a type II OPO, these twin beams are orthogonally polarized. It is thus easy to separate them spatially and to subsequently make their polarizations parallel. Then, the twin beams can be made indistinguishable by locking their frequency difference to zero, which has been an experimental challenge. This Letter reports the first experimental demonstration of nonclassical interference of such macroscopic boson fields. 

\begin{figure}[htb]
\begin{center}
\begin{tabular}{c}
\includegraphics[height= 1in]{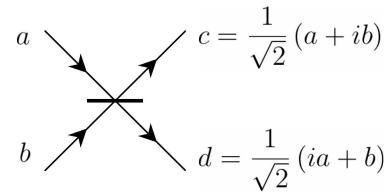}
\end{tabular}
\end{center}
\caption{Beam splitter. The input and output modes have annihilation operators ($a$,$b$) and ($c$,$d$) respectively. The reflection/transmission coefficients are $r=i t=2^{-1/2}$. The beams are aligned so that $\hat k_{c,d}=\hat k_{b,a}$.}
\label{bs}
\end{figure}

In general, a quantum interference experiment consists in ``splitting" a quantum field into two subfields, each experiencing its own phase evolution, and then ``recombining" these subfields and performing a measurement. The corresponding probability distribution presents fringes which give information about the phase difference of the two subfields. Examples include the Mach-Zehnder interferometer for light and the Ramsey interferometer for matter, which are isomorphic to each other. ``Nonclassical interference" may either mean that waves of a nonclassical nature are involved (e.g.\ matter waves), or that their behavior itself has no classical optical equivalent. The latter situation is what interests us, and is determined by the role of the vacuum modes of the quantum field, i.e.\ the physics of the ``splitting." The physics of the beam splitter (Fig.~\ref{bs}) takes center stage here and also determines the phase measurement noise. 
Take the example of an initial $N$-photon Fock state $\sket{\hat k_a,\hat\epsilon,\omega; N}_a$, where $\hat k$ and $\hat\epsilon$ are the unit wave and polarization vectors and $\omega$ the angular frequency. The beam splitter output is given by the interference of this state with the corresponding polarization- and frequency-degenerate vacuum state $\sket{\hat k_b,\hat\epsilon,\omega; 0}_b$. As was first demonstrated by Caves in 1980 \cite{caves80}, this yields a (classically intuitive) binomial probability distribution of the photon number between modes $c$ and $d$, with standard deviation $\Delta N_-^{out}=\Delta(N_c-N_d)\propto N^{1/2}$. Using the Heisenberg inequality between number and phase differences $\Delta N_-\Delta\phi_-\geq 1$, we obtain $\Delta\phi_-^{out}\propto N^{-1/2}$, which we call the classical limit of an interferometer. Note that this limit becomes $\langle N\rangle^{-1/2}$ for any input state of average photon number $\langle N\rangle$ and is independent of the photon statistics of the input \cite{caves80}. The classical limit is the limit of all interferometers whose input splitting involves unmodified vacuum modes, including last-generation atomic clocks \cite{clairon}. It is not, however, the ultimate phase detection limit, which can be obtained from a heuristic argument \cite{ou97}: by maximizing $\Delta N_-^{out}$ in a minimum uncertainty state, one gets $\Delta N_-^{out}\sim \langle N\rangle  \Leftrightarrow \Delta\phi_-^{out} \sim \langle N\rangle^{-1}$, called the Heisenberg limit. The physical picture here is that boson indistinguishability between the interferometer subfields extends to the total particle number.

It is well known that the key to reaching the Heisenberg limit is to modify the complementary input $b$ of the beam splitter.  The first proposal was to use a squeezed vacuum state \cite{caves81}, which was realized experimentally by Xiao {\em {\em et al.}}\  and Grangier {\em {\em et al.}}\ \cite{hli87}. A number of other proposals have been made. One of the simplest ones, from Holland and Burnett \cite{holland}, is to use number-correlated, degenerate input states of the general (ideal) density matrix 
\begin{equation}
\sum_{n,p}\rho_{np} \sket{\hat k_a,\hat\epsilon,\omega; n}\sket{\hat k_b,\hat\epsilon,\omega; n}\sbra{\hat k_a,\hat\epsilon,\omega; p}\sbra{\hat k_b,\hat\epsilon,\omega; p},
\label{rho}
\end{equation} 
which we will call twin modes. Such states give the Heisenberg limit as well \cite{kim}, even though the measurement procedure is complicated by the fact that the output intensities show no interference fringes. A Bayesian detection procedure \cite{holland} was proposed to remedy the situation and we have confirmed its experimental feasibility by numerically simulating nonideal conditions \cite{pooser}. 

A crucial requirement for the performance of twin-mode interferometry is the exact polarization and frequency degeneracy, i.e.\ indistinguishability, of the input states. This is the key to the nonclassical interference that maximizes $\Delta N_-^{out}$ and therefore minimizes $\Delta\phi_-^{out}$. A good illustration of this point is the simplest possible case of a twin photon pair $\sket{\hat k_a,\hat\epsilon,\omega; 1}_a\sket{\hat k_b,\hat\epsilon,\omega; 1}_b$, i.e.\ the \hom (HOM) interferometer \cite{hom}, recently revisited by Santori {\em et al.}\ using two consecutive photons from the same source \cite{santori}. Out of the four possible beam-splitter scattering probability amplitudes, the two corresponding to the output state  $\sket{\hat k_c,\hat\epsilon,\omega; 1}_c\sket{\hat k_d,\hat\epsilon,\omega; 1}_d$ interfere destructively because of indistinguishability of the input photons, i.e.\ the degeneracy of the input modes. This maximizes $\Delta N_-$, giving $\Delta N_-=2$ instead of $\sqrt 2$. This effect disappears if the beam splitter is misaligned ($\hat k_{c,d}\neq\hat k_{b,a}$), or if $\hat\epsilon_a\neq\hat\epsilon_b$ or $\omega_a\neq\omega_b$. If any of the previous conditions is true, for example $\omega_a\neq\omega_b$, then there are four input modes instead of two: the beam splitter input state becomes $\sket{\hat k_a,\hat\epsilon,\omega_a; 1}_a\sket{\hat k_a,\hat\epsilon,\omega_b; 0}_a\sket{\hat k_b,\hat\epsilon,\omega_a; 0}_b\sket{\hat k_b,\hat\epsilon,\omega_b; 1}_b$, and the photons only interfere with their respective vacuum modes instead of interfering with each other. One then obtains the probabilistic mixture of two single-photon cases: no destructive interference occurs and all possible output states are equiprobable.

Of even greater interest is the situation where the beam splitter input is $\sket{\hat k_a,\hat\epsilon,\omega; N}_a\sket{\hat k_b,\hat\epsilon,\omega; N}_b$ \cite{campos} or, better yet, the more general form expressed by Eq.~(\ref{rho}), for which the physics is identical \cite{kim}. Close approximations to such states are generated above threshold in a type II OPO, which emits intense, laser-like, cross-polarized, number-difference squeezed beams \cite{mertz91} that can be frequency-stabilized by use of standard techniques of laser metrology \cite{hall,wong93} or optical self-locking \cite{wong98}. However, squeezing and ultrastable frequency degeneracy have never been combined in the same experiment before, which is the {\em sine qua non} condition for the observation of the nonclassical twin-beam interference. We now present the experimental observation of this effect.

\begin{figure}[htb]
\begin{center}
\begin{tabular}{c}
\includegraphics[height=2in]{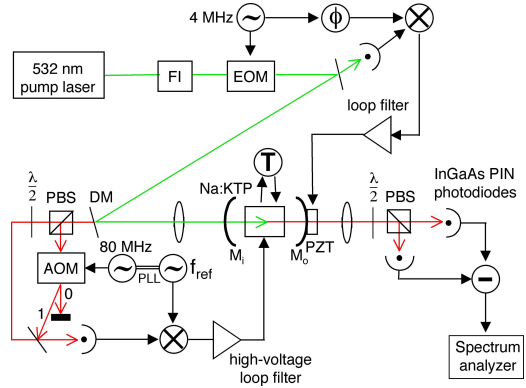}
\end{tabular}
\end{center}
\caption{Simplified experimental setup. Green lines denote the 532 nm pump beam. Red lines denote the 1064 nm OPO twin beams. The servo loop at the top of the figure is the optical frequency-lock loop, the one at the bottom left is the optical phase-lock loop. T: crystal temperature servo loop. $\rm M_i$: input mirror (reflectivity: $\simeq$ 0\% @ 532 nm; 99.99\% @ 1064nm). $\rm M_o$: output mirror (reflectivity: 99.995\% @ 532 nm; 99\% @ 1064nm). FI: Faraday isolator. EOM: electro-optic modulator. DM: dichroic mirror. PBS: polarizing beam splitter. AOM: acousto-optic modulator. PLL: (electronic) phase-lock loop (the 80 MHz and $\rm f_{ref}$ sources are electronically phase-locked together). PZT: piezoelectric transducer.}
\label{setup}
\end{figure}

\begin{figure}[htb]
\begin{center}
\begin{tabular}{c}
\includegraphics[height= 1.5in]{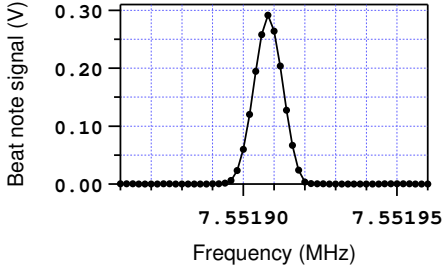}
\end{tabular}
\end{center}
\caption{Beat note signal of the phase-difference-locked OPO beams. Measurement bandwidth: 10 Hz. A frequency count of the peak by the spectrum analyzer yields no drift at 1 Hz resolution.}
\label{pll}
\end{figure}

The principal difficulty of the experiment resides in creating exactly frequency degenerate twin beams. A type-II OPO has a clustered emission spectrum (see \cite{feng} and Refs.\ therein) and the mode-hop length in our case is $\lambda/500$, compared to $\lambda/2$ for a laser. High performance servo loops are therefore essential to keep the OPO emission at the degenerate mode $\omega_a=\omega_b$, where $\omega_{a,b}$ are the twin beam frequencies. Our experimental setup is sketched in Fig.~\ref{setup} and is described in more detail in Refs.~\cite{feng,feng2}.
The OPO consists of a temperature-stabilized Na:KTP nonlinear crystal, in which pump photons at 532 nm are downconverted to cross-polarized pairs at 1064 nm, and an optical cavity formed by mirrors $\rm M_i$, $\rm M_o$. The 1064 nm twin beams exit through $\rm M_o$ to the right of the figure. The reflected pump beam is used for locking the cavity and a weak leak at 1064 nm through $\rm M_i$ is used for locking the phase difference of the twin beams. With only the temperature and cavity locks, the frequency difference error is $\pm$150 kHz \cite{feng}. Adding the optical phase-lock loop (PLL) reduces this error by more than 5 orders of magnitude to less than 1 Hz (Fig.~\ref{pll}),
while keeping the frequency difference continuously temperature-tunable over a few tens of MHz \cite{feng2}, a feature that optical self-phase-locking does not possess \cite{wong98}. Finally, since the PLL error signal is obtained from beams leaking through a mirror with 0.01\% transmission, it is entirely classical and the PLL cannot modify the quantum phase fluctuations, to which we now turn. 

Outside $M_o$, the half-wave plate and polarizer assembly behaves like the polarizer alone when the wave plate's axes are aligned with those of the polarizer and of the OPO crystal, and like a balanced beam splitter when the wave plate's axes are at a $\theta=\pi/8$ angle from those of the polarizer and of the crystal. We define the variables  $N_k=\d kk$, $X_k=(k+\d k)/\sqrt 2$, $P_k=i(\d k-k)/\sqrt 2$, $N_-^{in} = N_a-N_b$, and $N_-^{out}(\theta) = N_c(\theta)-N_d(\theta)$. Thus, $N_-^{out}(0)=N_-^{in}$, the twin-beam intensity difference, and $N_-^{out}(\pi/8)=i(\d ab-a\d b) = X_aP_b-X_bP_a$, the twin-beam interference term. Linearizing the small quantum fluctuations, we write field operators as $X=\langle X\rangle+\delta X=x+\delta X$. We then obtain the standard deviations
\begin{eqnarray}
&\Delta N_-^{out}(0)  =  |x| \Delta(\delta X_a-\delta X_b) = & |x| \Delta X_-^{in}\\
&\Delta N_-^{out}(\pi/8)  =  |x| \Delta(\delta P_a-\delta P_b) = & |x| \Delta P_-^{in},\label{degphase}
\end{eqnarray}
where Eq.~(\ref{degphase}) is only valid for indistinguishable twin beams. Thus, measuring $\Delta N_-$ before and after the beam splitter allows us to measure the two conjugate quadrature differences of the OPO and to test the Heisenberg inequality $\Delta X_-\Delta P_-\geq 1$. It is well known that the nondegenerate OPO gives amplitude quadrature correlations, i.e.\ squeezes $X_-$ (and $N_-$), and phase quadrature anticorrelations, i.e.\ antisqueezes $P_-$ (and $\phi_-$). Theoretical predictions have been given based on a semiclassical theory \cite{josab87,jdp} and on a fully quantum analysis \cite{reid,reid2}. The twin-beam intensity-difference spectra before and after the beam splitter are, respectively \cite{jdp}
\begin{eqnarray}
S_{N_-(0)}^{out} (u) & = & S_0\, S_{X_-}^{in}(u)  =  S_0\, \left(1 - \frac{\xi}{1+u^2}\right)\label{nsq} \\
S_{N_-(\pi/8)}^{out} (u) & = & S_0\, S_{P_-}^{in}(u)  =  S_0\, \left(1 + \frac{\xi}{u^2}\right),\label{psq}
\end{eqnarray}
where $S_0$ is the total shot noise amplitude of both beams, $u=\nu/\delta$ is a normalized frequency, $\delta=(T+A)D/2\pi$ is the cold-cavity FWHM, $D$ is the free spectral range, $\xi=T/(T+A)$ is the correlation coefficient, $T$ is the output coupler transmissivity, and $A$ is the single-pass intensity loss. Equations (\ref{nsq},\ref{psq}) can also be understood qualitatively: for very short observation times compared to the cavity storage time ($u\gg 1$), the photon correlations are destroyed by the random cavity-exit times of each twin of a pair and no squeezing is present [$S_{N_-(0)}^{out} (u\gg 1)  =  S_0$]. However, if one integrates the photon counting over at least the cavity storage time ($u\leq 1$) then correlations become visible again and squeezing of $N_-$ is observed within the cavity linewidth [Eq.(\ref{nsq})]. Hence, the conjugate variable $\phi_-$ should be antisqueezed within the cavity linewidth [Eq.(\ref{psq})], where the OPO's double resonance condition is relaxed, since the phase difference is fixed to a multiple of $2\pi$ by double resonance only for times not exceeding the cavity storage time ($u\geq 1$). For longer times ($u\ll 1$), the phase difference noise becomes dominated by the Schawlow-Townes drift of the OPO phase difference \cite{graham}. Record amounts of number-difference squeezing (-8.5 dB) have been obtained by the group of Fabre and Giacobino \cite{mertz91}, but the demonstration of phase-difference anti-squeezing  (\ref{psq}), which necessitates twin beam indistinguishability, has not been previously achieved. 
\begin{figure}[htb]
\begin{center}
\begin{tabular}{ll}
\includegraphics[height=2.4 in]{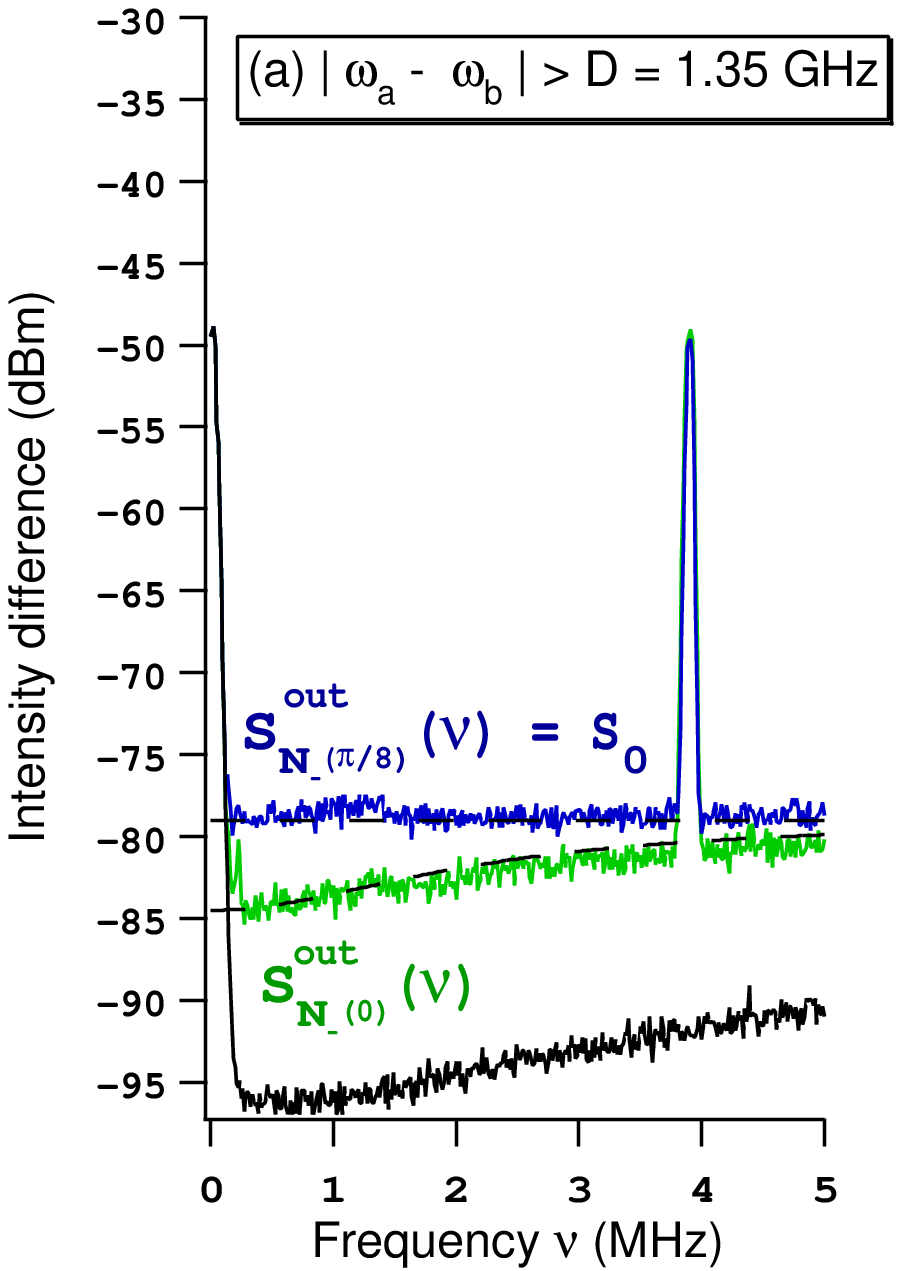}&
\includegraphics[height=2.4 in]{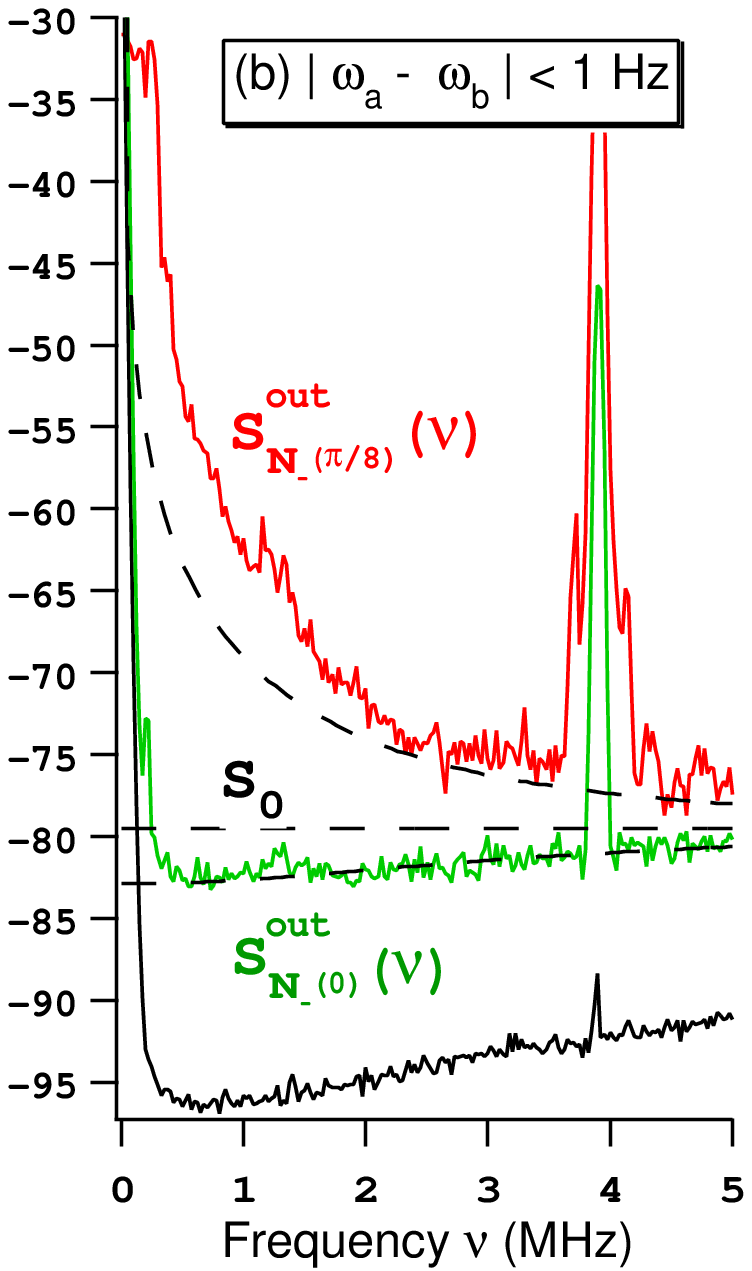}
\end{tabular}
\end{center}
\caption{Intensity-difference spectra for (a) distinguishable and (b) indistinguishable twin beams. Measurement bandwidth: 30 kHz. Green lines are intensity differences before the beam splitter ($\theta=0$), blue and red lines after ($\theta=\pi/8$). Bottom black lines are electronic detection noise floors. Black dashed lines are theoretical fits, of parameters ($S_0$, $\xi$, $\delta$): ($-79$ dBm, $0.72$, $2.98$ MHz) for (a) and ($-79.5$ dBm, $0.5$, $4.3$ MHz)  for (b). The 3.9 MHz peak is EO modulation. }
\label{res}
\end{figure} 

Figures \ref{res}(a,b) display the intensity difference spectra for distinguishable and undistinguishable OPO beams, respectively. In Fig.\ \ref{res}(a) the OPO frequencies $\omega_{a,b}$ are separated by at least the free spectral range, i.e.\ are totally distinguishable. The phase difference noise spectrum (blue line) is a measure of the total shot noise, each beam interfering with vacuum in the other input port. No HOM-type quantum interference takes place. In Fig.\ \ref{res}(b), the OPO frequencies are phase-locked within less than 1 Hz of each other. In this case, the twin beams are totally indistinguishable and the phase difference spectrum (red line) becomes extremely noisy for $\nu\leq\delta$. This noise is well fitted by Eq.~(\ref{psq}) despite additional technical noise (mainly pump intensity noise) at frequencies below 2 MHz. The fit parameters are entirely determined by the fit of the {\em intensity} difference spectrum (green line). This good agreement indicates that the OPO output is in a minimum uncertainty state. The intensity-difference noise is not affected by indistinguishability, even though we obtain different squeezing levels: - 5.5 dB (-6 dB when accounting for detection noise) in Fig.\ \ref{res}(a) but  only -3 dB in Fig.\ \ref{res}(b). We believe that this is due, in the latter case, to partial overlap of the beams with grey tracks (optical damage) in the crystal, which induces differential losses that cannot be balanced out. This is consistent with the slight increases of $\delta$ and of residual classical noise in Fig.~\ref{res}(b), compared to Fig.~\ref{res}(a). 

In conclusion, we have observed, for the first time, the HOM interference of macroscopic boson modes. Although their quantum state is much more complicated than, say, a twin Fock state, nonclassical interference does take place nonetheless, as was predicted in \cite{reid,jdp,kim}. The independence of the quantum interference from {\em common-mode} photon statistics stems from beam splitter physics, not from the light source itself. This is identical to the independence of the classical limit from the statistics of the source originally proven by Caves \cite{caves80}. Considering this result from the view point of the quantum variables defined {\em after} the beam splitter, we expect a minimum uncertainty state antisqueezed in $N_-$ and squeezed in $\phi_-$, from the beam splitter's property of swapping number- and phase-difference fluctuations \cite{yurke,holland}. Finally, the level of performance of this experiment is precisely that which is required in order to test the proposal of Reid and Drummond to create bright ultrastable EPR beams \cite{reid}. Because the OPO frequency difference is tunable in our experiment, we expect to be able to entangle frequency nondegenerate fields.

\begin{acknowledgments}
We thank Ken Nelson and Harvey Sugerman for their help with electronics design and construction, and Dan Perlov from Coherent-Crystal Associates for help with understanding Na:KTP. This work is supported by ARO grants DAAD19-01-1-0721 and DAAD19-02-1-0104, and in part by NSF grants PHY-0240532 and EIA-0323623.
\end{acknowledgments}

\end{document}